\documentclass{sig-alternate-05-2015}
\usepackage{multirow}

\begin{document}

\conferenceinfo{KDD 2016 Fashion Workshop,}{August 14, 2016, San Francisco, CA, USA}

\setcopyright{rightsretained}

%
%

\title{Fashion DNA: Merging Content and Sales Data for Recommendation and Article Mapping}
\numberofauthors{1}
\author{
%
%
\alignauthor
Christian Bracher, Sebastian Heinz, Roland Vollgraf
\titlenote{postal address of all authors: Zalando~SE, Charlottenstr.~4, 10969~Berlin, Germany} \\[1mm]
       \affaddr{Zalando Research}\\[1mm]
       \email{\{christian.bracher, sebastian.heinz, roland.vollgraf\}@zalando.de}
}
\maketitle
\begin{abstract}
We present a method to determine {\sl Fashion DNA}, coordinate vectors
locating fashion items in an abstract space.  Our approach is based
on a deep neural network architecture that ingests curated article information
such as tags and images, and is trained to predict sales for a large set of 
frequent customers. In the process, a dual space of customer style preferences 
naturally arises. Interpretation of the metric of these spaces is 
straightforward: The product of Fashion DNA and customer style vectors yields 
the forecast purchase likelihood for the customer--item pair, while the angle
between Fashion DNA vectors is a measure of item similarity.
Importantly, our models are able to generate unbiased purchase probabilities for 
fashion items based solely on article information, even in absence of sales
data, thus circumventing the ``cold--start problem'' of collaborative recommendation approaches.  Likewise, it generalizes 
easily and reliably to customers outside the training set.
We experiment with Fashion DNA models based on visual and/or tag item data, 
evaluate their recommendation power, and discuss the resulting article
similarities.
\end{abstract}
%
%
%
%

%
%


\section{Introduction}

Compared to other commodities, fashion articles are tough to characterize:  They 
are extremely varied in function and appearance and lack standardization, and are 
described by a virtually open-ended set of properties and attributes, including labels 
such as brand (``Levi's'') and silhouette (``jeans''), physical properties like shape
(``slim cut''), color (``blue''), and material (``stonewashed cotton denim''), 
target groups (``adult,'' ``male''), price, imagery (photos, videos), and customer and 
expert sentiments.  Many of these attributes are subjective 
or evolve over time; others are highly variable among customers, such as fit and style.

Zalando is Europe's leading online fashion platform, operating in 15 countries, with a base of $> 10^7$ customers and a catalog
comprising $> 10^6$ articles (SKUs), with $> 10^5$ SKUs available for sale online at any 
given moment.  Given the incomprehensibly large catalog and the heterogeneity of the 
customer base, matching shoppers and articles requires an automated pre-selection 
process that however has to be personalized to conform to the customers' individual 
styles and preferences:  The degree to which items are similar or complement each other
is largely in the eye of the beholder.  At the same time, proper planning at a complex 
organization like Zalando requires an ``objective'' description of each fashion 
item -- one that is traditionally given by a curated set of labels like the 
above-mentioned. Still, these ``expert labels'' are often ambiguous, cannot 
consider the variety of opinion found among shoppers, and often do not even match the 
consensus amongst customers.

We set ourselves the task to find a mathematical representation of fashion items
that permits a notion of similarity and distance based on the expert labels, as well as 
visual and other information, but that is designed to maximize sales by enabling tailored
recommendations to individual clients. Our vehicle is a deep feedforward neural 
network that is trained to forecast the individual purchases of a sizable number 
$(10^4 \ldots 10^5)$ of frequent Zalando customers, currently using expert labels and 
catalog images for $\sim 10^6$ past and active Zalando articles as input.
For each SKU we compute the activation of the topmost hidden layer in the network which 
we call its {\sl Fashion DNA} (fDNA).
The idiosyncratic style and taste of our shoppers is then simultaneously encoded as a 
vector of equal dimension in the weight matrix of the output layer.  
Inner products between vectors in these two dual spaces yield a likelihood of purchase and also express
the similarity of fashion articles.
Section~\ref{kdd:network} explains the mathematical model and presents the architecture of our network.

Collaborative-filtering based recommender systems \cite{kdd:goldberg1992using,kdd:BellKor} suffer from the ``cold-start problem'' \cite{kdd:coldstart}:
Individual predictions for new customers and articles cannot be calculated in the absence of prior purchase data.
Training of our network relies on sales records \textit{and} content-based information on the article side, while only purchase data is used for customers. Once trained, the model computes fDNA vectors based on content article data alone, thus alleviating the ``cold-start problem.'' 
Our experiments show that the quality of such recommendations is only moderately 
affected by the lack of sales data. Likewise, it is easy to establish style vectors
for unseen customers (by standard logistic regression), and their 
forecast performance is almost indistinguishable from training set customers.
The generalization performance of our Fashion DNA model is demonstrated in Section~\ref{kdd:reco}.

As Fashion DNA combines item properties and customer sales behavior, distances in the
fDNA space provide a natural measure of article similarity.  In Section~\ref{kdd:similarity},
we examine the traits shared by neighboring SKUs, and use dimensionality reduction to unveil a
instructive, surprisingly intricate landscape of the fashion universe.

\section{The Fashion DNA Model}
\label{kdd:network}

We now proceed to describe the design of the network that yields the Fashion DNA. It
ingests images and/or attribute information about articles, and tries to predict
their likelihood of purchase for a group of frequent customers. After training, 
item fDNA is extracted from the internal activation of the network.

\subsection{Input data}
\label{kdd:network:data}

For our experiments, we selected a ``standard'' SKU set of 1.33 million items 
that were offered for sale at Zalando in the range 2011--2016, mostly
clothing, shoes, and accessories, but also cosmetics and housewares.  For 
each of these articles, one JPEG~image of size $177\,\times\,256$ was available
(some article images had to be resized to fit this format).  
Most of these images depict the article on a neutral
white background; a few were photographed worn by a model.

We furthermore used up to six ``expert labels'' to describe each item.  
These tags were selected to represent distinct qualities of the article 
under consideration:  Besides the brand and main color, they comprise the
commodity group (encoding silhouette, gender, age, and function), pattern,
and labels for the retail price and fabrics composition. Price labels were 
created using $k$--means clustering \cite{kdd:elements} of the logarithm of 
manufacturer suggested 
prices (MSRPs). For the fabrics composition labels, a linear space spanned 
by possible combinations of about 40 different fibers (e.~g., cotton and wool) 
was segmented into 80 clusters. Labels were issued only where applicable; for instance, 
fabric clusters are restricted to textile items. Generally, 
only labels with a minimum number of 50 SKUs were
retained. The number of 
different classes per label, and the percentage of articles tagged, are shown
in Table~\ref{kdd:network:tags}.
\begin{table}[b]
\centering
\begin{tabular}{|l|r|r|} \hline
Tag Description&Class Count&Coverage\\ \hline
Brand code & 2,401 & 97.2\%\\ \hline
Commodity group& 1,224 & 98.6\%\\ \hline
Main color code & 75 & 99.3\%\\ \hline
Pattern code & 47 & 32.6\%\\ \hline
Price cluster & 28 & 100.0\% \\ \hline
Fabric cluster & 80 & 64.9\%\\ 
\hline\end{tabular}
\caption{Article tags used for network training}\label{kdd:network:tags}
\end{table}
Label information was supplied to the network in the form of one-hot encoded
vectors of combined length 3,855. Missing tags were replaced by zero vectors.

For validation purposes, the item set was randomly split 9:1 
into a training subset $\mathcal{I}_\mathrm{t}$ of $N_\mathrm{t}\sim 1.2\text{m}$ and a validation subset $\mathcal{I}_\mathrm{v}$ of $N_\mathrm{v}\sim 130\text{k}$ articles.

\subsection{Purchase matrix}
\label{kdd:network:labels}

The Fashion DNA model is trained and evaluated on purchase information which
we supply in the form of a sparse boolean matrix $\Pi$, indicating the articles
that have been bought by a group of 30,000 individuals that are 
among Zalando's top customers. Their likelihood of purchase, averaged over
all customer--SKU combinations, was $p_\mathrm{avg}\,=\,1.14\cdot10^{-4}$, 
amounting to 150 orders per customer in the standard SKU set. We task 
the network to forecast these purchases by assigning a likelihood for 
a ``match'' between article and customer.

As we are interested in the generalization of these predictions to the many
Zalando customers not included in the selection, we likewise split the base
of 30,000 customers 9:1 into a training set $\mathcal{J}_\mathrm{t}$ of $K_\mathrm{t} = 27,000$ customers
and a validation set $\mathcal{J}_\mathrm{v}$ of $K_\mathrm{v} = 3,000$ customers, 
where care has been taken that the purchase frequency distributions in the two sets are aligned.

\begin{figure}
\centering
\includegraphics[width=0.5\columnwidth]{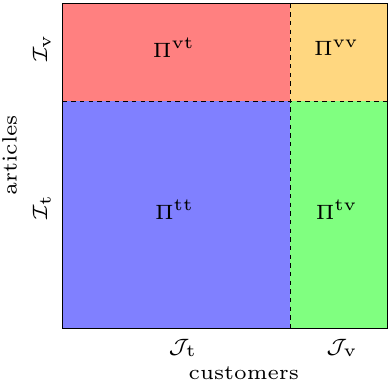}
\caption{Illustration of the purchase matrix $\Pi$. 
Every entry $\Pi_{ij}\in\{0,1\}$ encodes whether customer $j$ has bought article $i$. 
Since we split the articles as well as the customers into a validation and a training set, 
the purchase matrix $\Pi$ consists of four parts.}
\label{kdd:network:maths:fig1}
\end{figure}

Hence, the purchase matrix is made of 4 parts, compare Figure~\ref{kdd:network:maths:fig1}:
\begin{displaymath}
\begin{array}{ll}
\text{training data} & \Pi^{\mathrm{tt}}=\big(\Pi_{i,j}\big)_{i\in\mathcal{I}_\mathrm{t},j\in\mathcal{J}_\mathrm{t}},\\
\text{SKU validation data} & \Pi^{\mathrm{vt}}=\big(\Pi_{i,j}\big)_{i\in\mathcal{I}_\mathrm{v},j\in\mathcal{J}_\mathrm{t}},\\
\text{customer validation data} & \Pi^{\mathrm{tv}}=\big(\Pi_{i,j}\big)_{i\in\mathcal{I}_\mathrm{t},j\in\mathcal{J}_\mathrm{v}},\\
\text{SKU-customer validation data} & \Pi^{\mathrm{vv}}=\big(\Pi_{i,j}\big)_{i\in\mathcal{I}_\mathrm{v},j\in\mathcal{J}_\mathrm{v}}.\\
\end{array}
\end{displaymath}
We are going to deal with each of these parts separately.

\subsection{Mathematical model}
\label{kdd:network:maths}

A possible strategy to solve a recommender problem is by logistic factorization 
\cite{kdd:logisticfactorization}, \cite{kdd:matrixfactorization} 
of the purchase matrix $\Pi$. Following that strategy, 
the probability $p_{ij}$ is defined by
\begin{equation}
\label{kdd:eq:logreg}
p_{ij} \,=\, P(\Pi_{ij}=1)\,=\, \sigma\left( \mathbf{f}_i{\cdot} \mathbf{w}_j + b_j \right) \;,
\end{equation}
where $\sigma(x) = [1 + \exp(-x)]^{-1}$ is the logistic function. 
Moreover, $\mathbf{f}_i$ is a factor associated with SKU $i$, $\mathbf{w}_j$ a factor 
and $b_j$ a scalar associated with customer $j$. 
During training, when we are dealing with the segment $\Pi^{\mathrm{tt}}$ of the purchase matrix, 
the quantities $\mathbf{f}_i$, $\mathbf{w}_j$ and $b_j$ are adjusted such that the (mean) cross entropy loss
\begin{equation}
\label{kdd:eq:loss}
{\cal L}_\mathrm{tt} \,=\, 
{-}\frac{1}{N_\mathrm{t} K_\mathrm{t}} \,\sum_{i\in\mathcal{I}_\mathrm{t}}
\sum_{j\in\mathcal{J}_\mathrm{t}}
\left[ \Pi_{ij} \log p_{ij} + \left(1 {-} \Pi_{ij} \right) \log\left(1 {-} p_{ij}\right) \right]
\end{equation}
gets minimized. We define cross entropy losses on the remaining parts $\Pi^{\mathrm{vt}}$, 
$\Pi^{\mathrm{tv}}$ and $\Pi^{\mathrm{vv}}$ by changing the index sets in \eqref{kdd:eq:loss} appropriately.

Our method is based on Equations~\eqref{kdd:eq:logreg} and \eqref{kdd:eq:loss}, yet, 
it yields a ``conditioned factorization'' only. Instead of learning the factor 
$\mathbf{f}_i$ directly from the purchase data, we derive $\mathbf{f}_i$ in a deterministic way 
via the SKU feature mapping
\begin{equation}
\label{kdd:eq:DNN}
\mathbf{f}_i \, = \, f(\mathbf{\phi}_i;\theta),
\end{equation}
where $\mathbf{\phi}_i$ collects article features and metadata for SKU $i$. 
Moreover, $f$ is a family of nonlinear functions given by a deep neural network (DNN), 
and $\theta$ are parameters which are learned by our model. 
We remark that ``deep structured semantic models'' \cite{kdd:Huang_DSSM} 
solve problems similar to the one studied here. 

For every article $i$ let us identify $\mathbf{f}_i$ as its fDNA. 
Likewise, we can think of $\mathbf{w}_j$ as the ``style DNA'' reflecting the individual 
preferences of customer $j$. The bias $b_j$ is a measure for the overall propensity of customer $j$ to buy.
\begin{figure}
\centering
\includegraphics[width=\columnwidth]{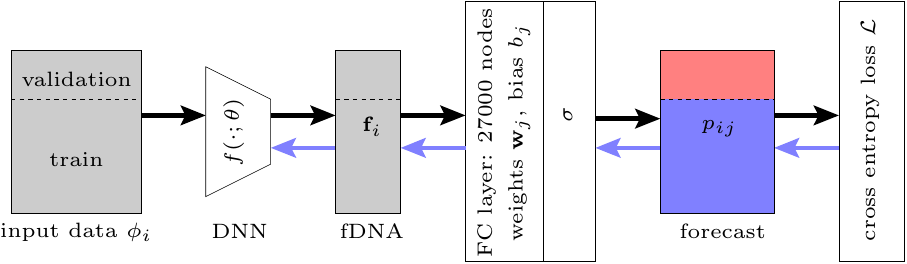}
\caption{Schematic network architecture for Fashion DNA retrieval: training set customers. Article
information (left) is passed through a deep neural network (DNN) to generate fDNA. The output is combined with customer-specific weights $\mathbf{w}_j$ and biases $b_j$
in a fully connected layer and transformed into customer--item pair 
purchase probabilities by logistic regression 
(Eq.~(\ref{kdd:eq:logreg})). These forecasts are
compared to the actual purchase matrix $\Pi$ (ground truth) in a
cross-entropy loss layer (Eq.~(\ref{kdd:eq:loss})). The 
loss for the training set of articles is minimized by
backpropagation (indicated by blue arrows).}\label{kdd:network:architecture:fig1a}
\end{figure}

In addition to training, there are three validation procedures:
\begin{enumerate}
\item\label{kdd:enum:training} 
\textbf{Training:} The purchase data in $\Pi^{\mathrm{tt}}$ is used to train the 
network parameters $\mathbf{\theta}$ and the customer weights $\mathbf{w}_j$ and 
biases $b_j$, $j\in\mathcal{J}_\mathrm{t}$, in order to perform the logistic matrix 
``conditioned factorization'' via \eqref{kdd:eq:logreg}, as sketched in 
Figure~\ref{kdd:network:architecture:fig1a}.
The loss is given by Equation~\eqref{kdd:eq:loss}.
\item\label{kdd:enum:sku_validation} 
\textbf{SKU validation:} For SKUs not seen during training, $\Pi^{\mathrm{vt}}$ 
contains purchases of the training customers. Validation is straightforward:
Part~\ref{kdd:enum:training} yields the weights $\mathbf{w}_j$ and 
bias $b_j$ for a training customer $j\in\mathcal{J}_\mathrm{t}$, as well as
the parameters $\mathbf{\theta}$. We then compute the 
fDNA $\mathbf{f}_i$ for SKUs $i\in\mathcal{I}_\mathrm{v}$ via Equation~\eqref{kdd:eq:DNN}.
\item\label{kdd:enum:customer_validation} 
\textbf{Customer validation:} Validation of new customers only works indirectly. 
It requires additional training of parameters, i.~e., we have to find weights
$\mathbf{\mathbf{w}}_{j}$ and biases $b_j$ for the new customers $j\in\mathcal{J}_\mathrm{v}$. 
This amounts to $K_\mathrm{v}$ simultaneous logistic regression problems from
$\mathbf{f}_{i}$ to $\Pi_{ij}$, where the $\mathbf{f}_{i}$ are taken from 
Part~\ref{kdd:enum:training} (see Figure~\ref{kdd:network:architecture:fig1b}). 
The validation loss ${\cal L}_\mathrm{tv}$ measures the fDNA's ability to 
linearly predict purchase behavior for such customers. 
High values compared to the training loss ${\cal L}_\mathrm{tt}$
indicate that the SKU feature mapping $f(\cdot;\theta)$
does not generalize well to validation set customers.
\begin{figure}
\centering
\includegraphics[width=\columnwidth]{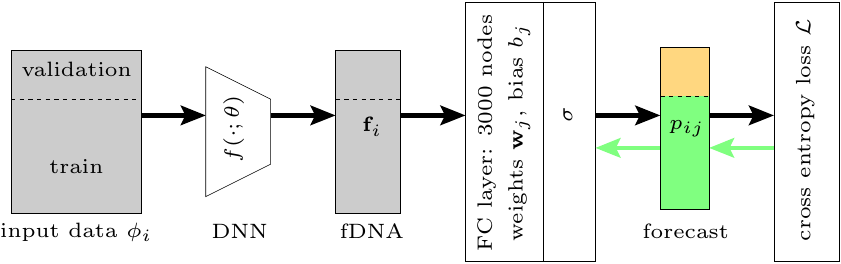}
\caption{Schematic network architecture for Fashion DNA retrieval: validation set customers. 
For a hold-out set of customers, backprop (green arrows) is
limited to the weights and biases, leaving the fDNA fixed. Evaluation
of performance mainly takes place on the combined validation set of
customers and items (orange).}\label{kdd:network:architecture:fig1b}
\end{figure}
\item\label{kdd:enum:customer_sku_validation} 
\textbf{SKU-customer validation:} Eventually, on $\Pi^{\mathrm{vv}}$, 
we measure whether the logistic regression learned during customer
validation in Part~\ref{kdd:enum:customer_validation} generalizes well
to unseen SKUs, analogous to the procedure in Part~\ref{kdd:enum:sku_validation}. 
The factors $\mathbf{w}_{j}$ now correspond to
the validation customers $j\in\mathcal{J}_\mathrm{v}$,
which were derived in Part~\ref{kdd:enum:customer_validation} 
by means of logistic regression.
\end{enumerate}

\subsection{Neural network architecture}
\label{kdd:network:architecture}

We experimented with three different multi-layer neural network models (DNNs)
that transform attribute and/or visual item information into fDNA. The width
of their output layers defines the dimension $d$ of the Fashion DNA space 
(in our case, $d=256$). This fDNA then enters the logistic regression layer
(Figures~\ref{kdd:network:architecture:fig1a} 
and~\ref{kdd:network:architecture:fig1b}).

\subsubsection{Attribute-based model}
\label{kdd:network:attribute}

Here, we used only the six attributes listed in Table~\ref{kdd:network:tags}
as input data.  As described in Section~\ref{kdd:network:data}, these
``expert'' tags were combined into a sparse one-hot encoded vector that was
supplied to a four-layer, fully connected deep neural network with steadily
diminishing layer size.  Activation was rendered nonlinear by standard ReLUs, 
and we used drop-out to address overfitting.  The output yields 
``attribute Fashion DNA'' based only on the six tags.

\subsubsection{Image-based model}
\label{kdd:network:image}

This model rests on purely visual information.  It feeds Zalando catalog images 
(see Section~\ref{kdd:network:data} for details) into an eight-layer convolutional 
neural network based on the AlexNet architecture \cite{kdd:alexnet},
resulting in ``image fDNA.''. 

\subsubsection{Combined model}
\label{kdd:network:combo}

\begin{figure}
\centering
\includegraphics[width=\columnwidth]{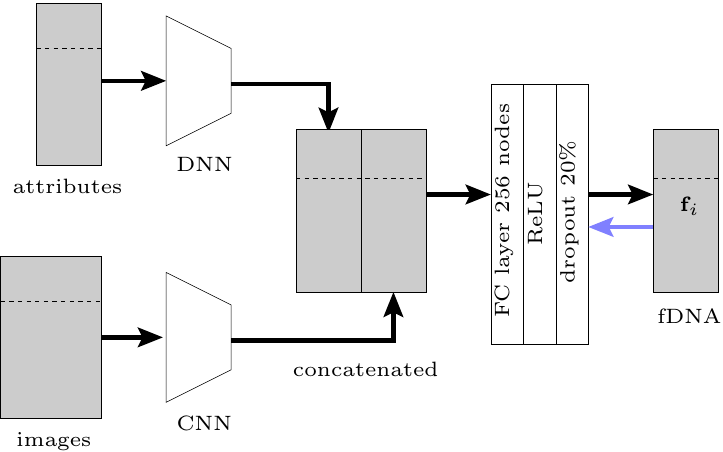}
\caption{Combining attribute- and image-based Fashion DNA.  Article data
flow through the respective pre-trained networks (see 
Sections~\ref{kdd:network:attribute} and~\ref{kdd:network:image});
the resulting attribute and image fDNAs are concatenated to a vector
of dimension $2d$ which is subsequently halved in size by a fully connected 
layer with ReLU nonlinearity.  Only the weights
in the merging layer are adjusted by backpropagation from the logistic
regression layer in Figures~\ref{kdd:network:architecture:fig1a}
and~\ref{kdd:network:architecture:fig1b}.}
\label{kdd:network:architecture:fig3}
\end{figure}

Finally, we integrated the attribute- and image-based architectures into a
combined model. We first trained both networks and then froze
their weights, so their outputs are the fDNA in either model. The
outputs are concatenated into a single input vector, then supplied to
a fully connected layer with adjustable weights that condenses 
this vector into the fDNA vector of the model.  This layer,
together with the logistic regressor, was trained by
backpropagation as described above.
The architecture is sketched in Figure~\ref{kdd:network:architecture:fig3}.

All three models were implemented in the Caffe framework \cite{kdd:caffe}
and used to generate fDNA vectors for the standard SKU set, with 
corresponding customer weights and biases. For training, the weights of 
the logistic regression layer are initialized with Gaussian noise, while
the bias is adjusted to reflect the individual purchase rate of customers on
the training SKU set.  The resulting fDNA vectors are sparse (about 30\%
non-zero components), leading to a compact representation of articles.

\section{FDNA-based recommendations}
\label{kdd:reco}

A key objective of Fashion DNA is to match articles with prospective
buyers.  Recall that the network yields a sale probability $p_{ij}$ for
every pair of article $i$ and customer $j$; it is therefore natural to
rank the probabilities by SKU for a given customer to provide a personalized
recommendation, or to order them by customer to find likely prospective
buyers for a fashion item. In this section, 
we examine the properties of fDNA-based item recommendations for customers.

\subsection{Probability distribution}
\label{kdd:reco:histogram}

Predicted probabilities span a surprisingly large range: The least likely
customer--item matches are assigned probabilities $p_{ij} < 10^{-12}$, too
low to confirm empirically, while the classifier can be extraordinarily
confident for likely pairings, with $p_{ij}$ approaching 50\%.  

To be valuable, quantitative recommendations need to be {\sl unbiased}:  
The predicted probability $p_{ij}$ should accurately reflect the observed 
chance of sale of the item to the customer.  As sales are binary 
events, comparing the probability with the ground truth requires aggregating 
many customer--article pairs with similar predicted likelihoods in order to 
evaluate the accuracy of the forecast.  As the average likelihood of an
article-customer match is small (on the order of only $10^{-4}$ even for 
the frequent shoppers included here), comparisons are affected by statistical 
noise, unless large sample sizes are used.  Fortunately, our model yields 
about $4\cdot10^9$ combinations of customers and fashion items in the 
respective validation sets.

\begin{figure}
\centering
\includegraphics[width=\columnwidth]{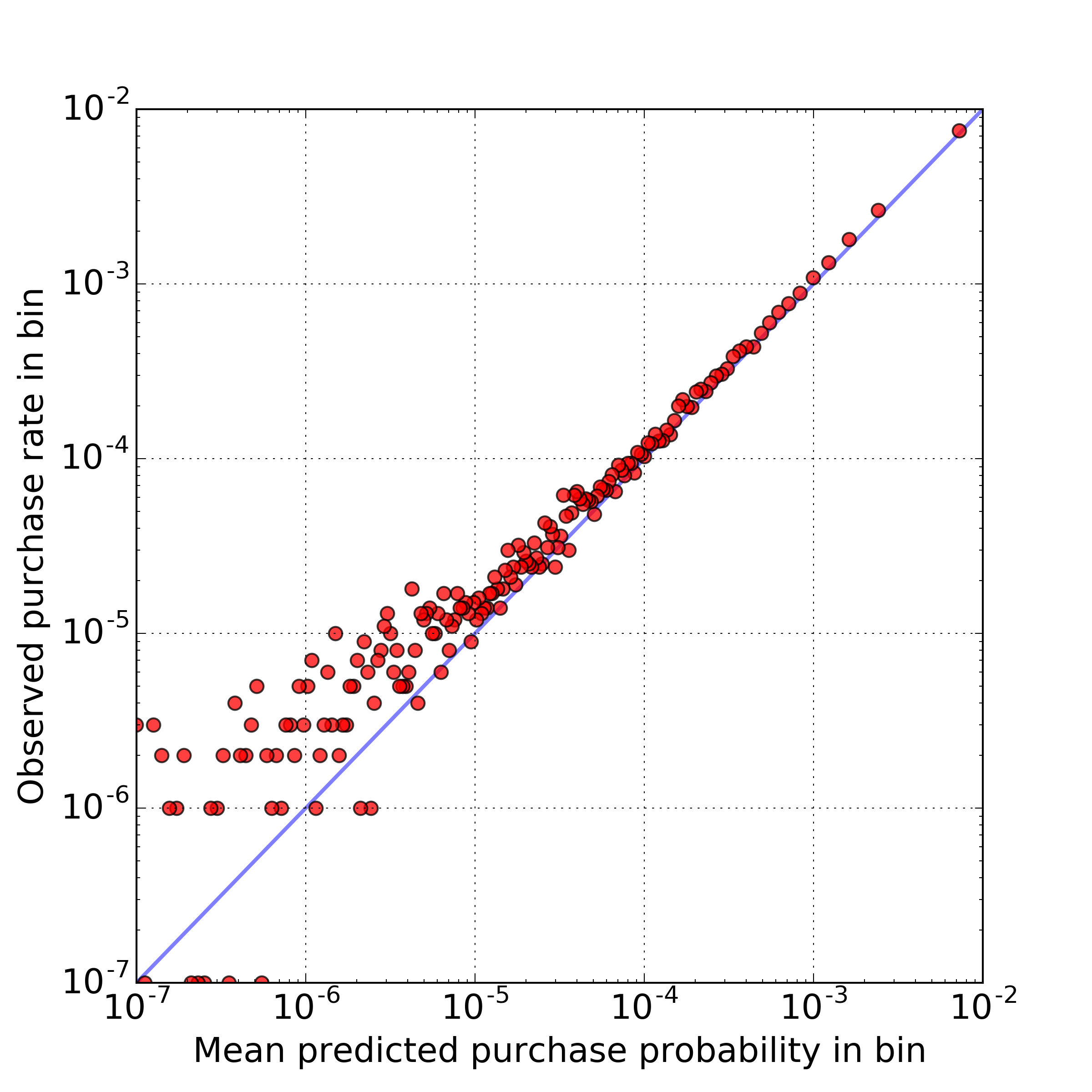}
\caption{Predicted probability vs.\ actual purchase rate for pairs in the 
customer--item validation set $(\Pi^\mathrm{vv})$ for the combined 
attribute--image network model. 
Each point represents a bin of $10^6$ pairs with similar predicted 
probability, taken from a random sample of $2\cdot10^8$ pairs.  
Bins with average probabilities $< 10^{-7}$ (representing $\sim\frac18$ of 
the sample) are not shown.}
\label{kdd:reco:histogram:fig2}
\end{figure}

An analysis of this type reveals that our models indeed are largely devoid
of such bias, as Figure~\ref{kdd:reco:histogram:fig2} illustrates. For the
analysis, $2\cdot10^8$ customer--item pairs have been sampled in the 
validation space, and then sorted into 200 bins by probability.  In the 
figure, the average probability of the $10^6$ members of each bin is
compared to the number of actual purchases per pair.  The predicted and
empirical rates track each other closely, save for pairs considered very
unlikely matches by the regressor. In this regime, the model underestimates 
the empirical purchase rate which settles to a base value of about
$1.5\cdot10^{-6}$.

\subsection{Recommendation quality}
\label{kdd:reco:quality}

In some contexts, quantitative knowledge of the predicted purchase likelihood
is less relevant than ranking a set of fashion articles by the preference of
individual customers. The overall quality of such a ranking can be 
illustrated by {\sl receiver operating characteristic} (ROC) analysis.

\subsubsection{ROC analysis}
\label{kdd:reco:roc}

An ideal model would rank those SKUs in the hold-out set highest that were
actually bought by the customer.  In reality, of course, the model will give
preference to some items not chosen by the customer, and the number of ``hits''
and ``misses'' will both grow with the number of recommendations issued, or,
equivalently, with a decreasing purchase probability threshold.
A common technique to analyze the performance of such a binary classifier 
with variable threshold is ROC analysis
\cite{kdd:elements}.  The ROC diagram displays the number of recommended sold
items (the ``hits,'' or in the parlance of ROC, the ``true positives''), as
a function of recommended, but not purchased items (the ``misses,'' or
``false positives'').  As the threshold probability is reduced from 1 to 0,
the true and false positive rates trace out a ROC curve leading from the
lower left to the upper right corner of the diagram.  As overall higher true 
positive rates indicate a more discriminating recommender, the area under 
the ROC curve ({\sl AUC score}) is a simple indicator of the relative 
performance of the underlying model \cite{kdd:auc}, with $\mathrm{AUC} = 1.0$
being ideal, whereas $\mathrm{AUC} = 0.5$ is no better than guessing 
(diagonal line in diagram).

In principle, ROC analysis yields an individual curve and AUC score for each
customer $j$, and will depend on the choice of fDNA model. To evaluate generalization 
performance, we aggregated numerous customer-article pairs ($2\cdot10^{8}$)
for each of the four training/validation combinations 
$\Pi^\mathrm{tt}$, $\Pi^\mathrm{vt}$, $\Pi^\mathrm{tv}$, $\Pi^\mathrm{vv}$, 
predicted purchase likelihood, and conducted 
a ``synthetic'' ROC analysis on this data that presents an average over many 
customers instead.  We repeated this calculation for each fDNA model.
The outcomes are listed in condensed form as average AUC scores in 
Table~\ref{kdd:reco:auc}.
\begin{table}
\centering
\begin{tabular}{|c||c|c||c|c|}
\hline
\multirow{2}{*}{AUC Score} & \multicolumn{2}{c||}{Training SKUs} & \multicolumn{2}{c|}{Validation SKUs} \\ \cline{2-5} 
                & $\Pi^\mathrm{tt}$ & $\Pi^\mathrm{tv}$ & $\Pi^\mathrm{vt}$ & $\Pi^\mathrm{vv}$ \\ \hline
by attribute & 0.944 & 0.940 & 0.915 & 0.916 \\ \hline
by image     & 0.941 & 0.937 & 0.877 & 0.877 \\ \hline
combined     & \textbf{0.967} & \textbf{0.963} & \textbf{0.933} & \textbf{0.932} \\ \hline
\end{tabular}
\caption{AUC scores by network model}\label{kdd:reco:auc}
\end{table}

\subsubsection{Generalization performance}
\label{kdd:reco:generalize}

Before we discuss model performance in detail, we examine to which extent 
information gleaned from the training of our models is transferable to fashion items 
(e.g., SKUs newly added to the catalog) and customers outside the training set.

Table~\ref{kdd:reco:auc} tells us that for a given fDNA model, the AUC score
is highest if both customers and articles are taken from the training sets
($\Pi^\mathrm{tt}$), diminishes as one of the groups is switched
to the validation set ($\Pi^\mathrm{vt}$, $\Pi^\mathrm{tv}$), and
becomes lowest when both customers and articles are taken from the hold-out
sets ($\Pi^\mathrm{vv}$), as expected. But the scores conspicuously pair up:  
For a given SKU set (training or validation), there is very little difference 
in AUC score between the training and validation set customers. This observation 
holds irrespective of the specific fDNA model employed. 

\begin{figure}
\centering
\includegraphics[width=\columnwidth]{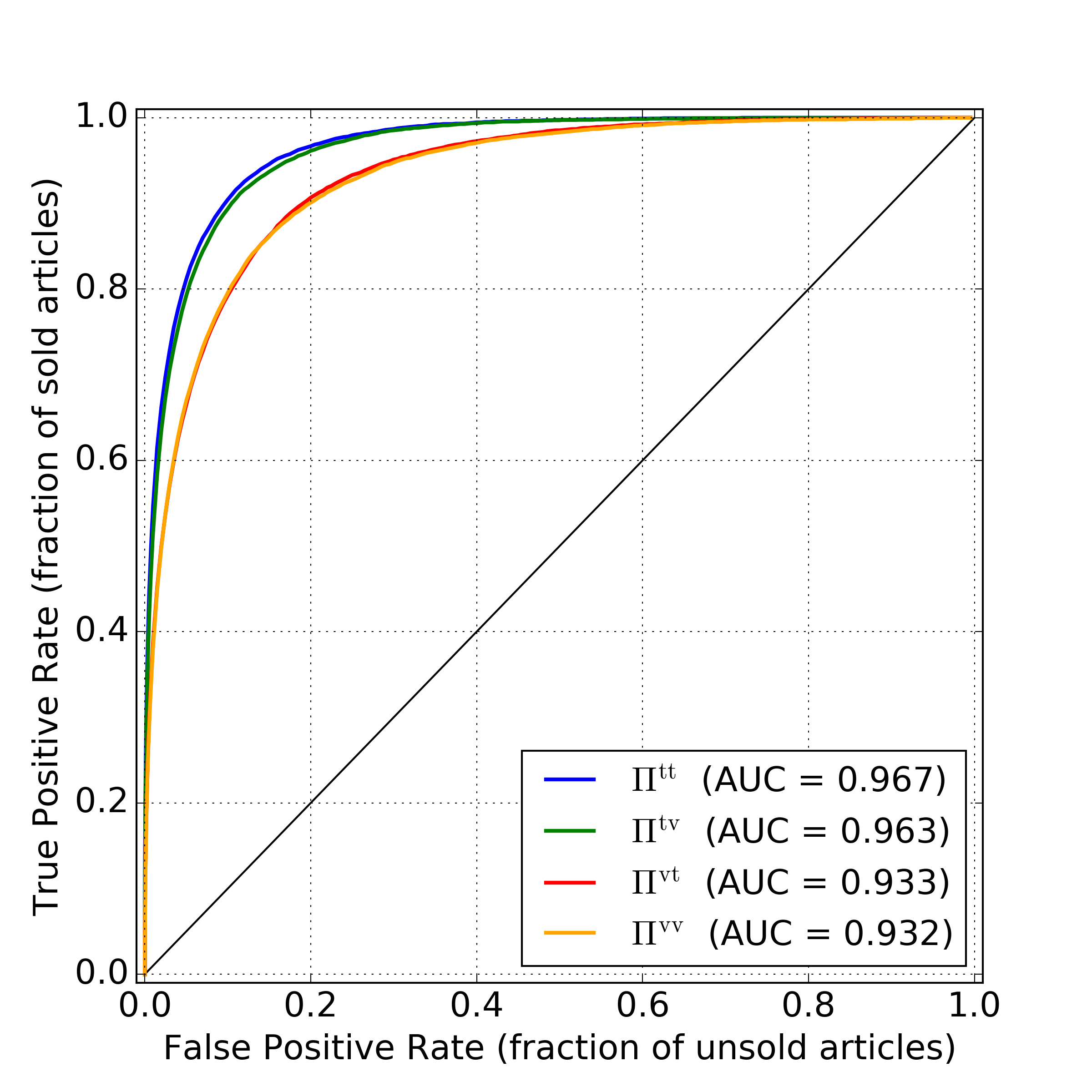}
\caption{Comparison of recommendation performance for training and validation set
items and customers in the combined attribute--image fDNA model.  The ROC 
curves for validation set customers (green, orange) closely follow the corresponding
curves for training set customers (blue, red).  There is a more 
pronounced difference between training set SKUs (blue, green) and hold-out
SKUs (red, orange). Color scheme as in Figure~\ref{kdd:network:maths:fig1}.}
\label{kdd:reco:roc:fig5}
\end{figure}

Picking the most powerful fDNA model based on images and attributes, we
determined ROC curves for the four cases laid out in Figure~\ref{kdd:network:maths:fig1}.  
Adopt the color code defined there, we display the curves in Figure~\ref{kdd:reco:roc:fig5}.
Indeed, the ROC curves for $\Pi^\mathrm{tt}$ (blue) and $\Pi^\mathrm{tv}$ (green)
track each other closely, as do the curves for validation set articles 
$\Pi^\mathrm{vt}$ (red) and $\Pi^\mathrm{vv}$ (orange).
We conclude that our approach generalizes extremely well to hold-out customers, 
at least when they shop at a similar rate.

Next, we address another related aspect of generalization performance for 
our model architecture:  How well does recommendation quality in the article 
training set correlate with the quality for items in the hold-out set for a
given customer?  Recommendation theory sometimes posits that most customers 
fall in well-defined groups that have aligned preferences, except for a few
``black sheep'' with idiosyncratic behavior that cannot be forecast well.
However, given the extreme amount of choice, and the sparsity of sales
coverage, it can be argued that idiosyncrasy in fashion taste and style is 
the norm rather than the exception.

\begin{figure}
\centering
\includegraphics[width=\columnwidth]{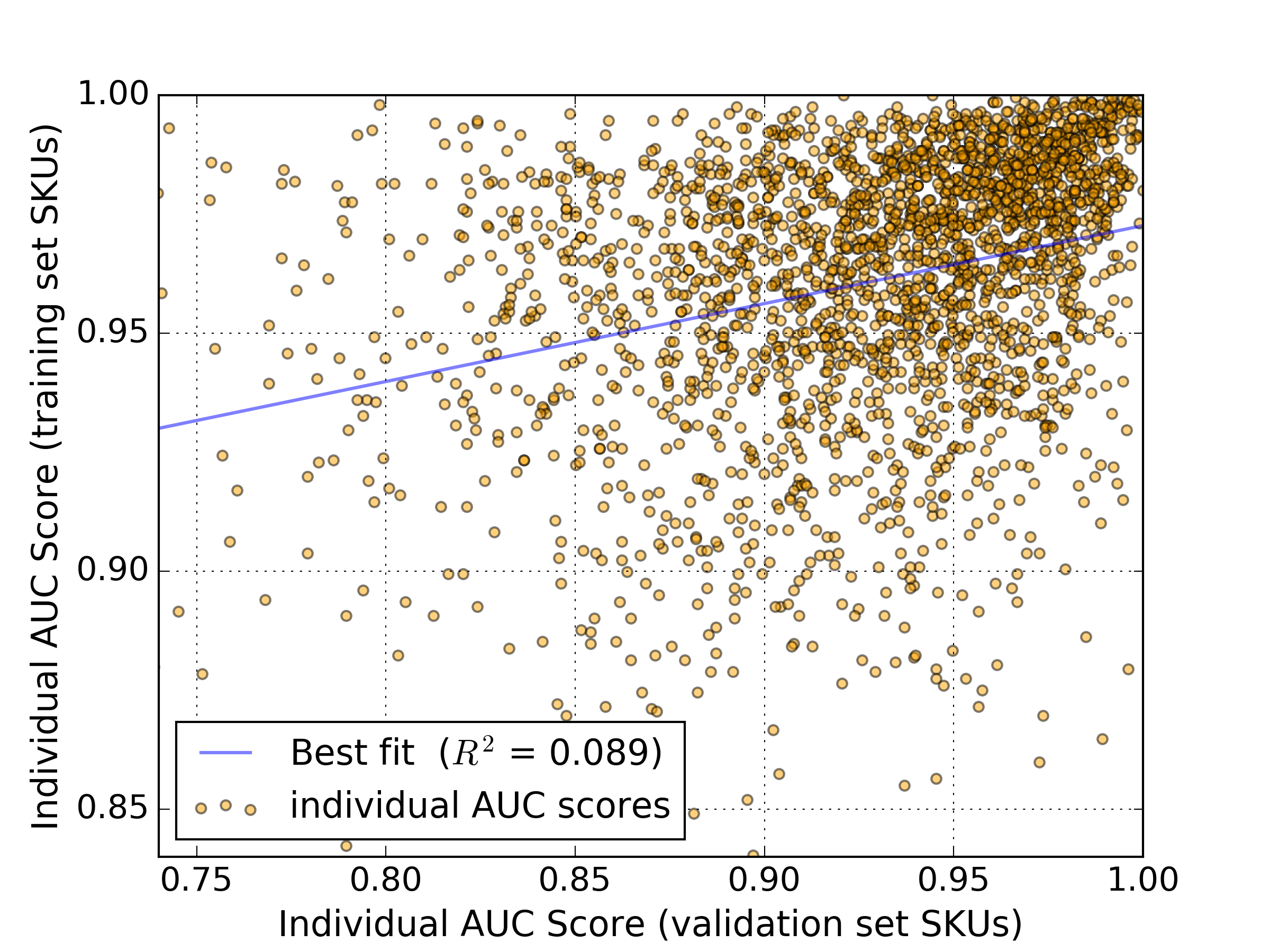}
\caption{Individual AUC scores in the combined fDNA model.  The plot compares 
recommendation performance for training and validation set SKUs for 3,000 customers
in the hold-out set.  Correlation of scores between the two sets is weak.}
\label{kdd:reco:roc:fig6}
\end{figure}

To investigate, we compiled recommendations for a large group of individual 
customers for both article training and validation sets, and compared them to their
actual purchases in both sets. Then, we examined the resulting pairs of 
customer-specific AUC scores for correlation.  
Figure~\ref{kdd:reco:roc:fig6} displays a 
scatterplot of such scores for the combined attribute-image fDNA model, using 
the hold-out customer set.  Although regression analysis reveals a weak
dependency (Pearson coefficient of correlation $R^2 \approx 0.09$), the plot is 
dominated by random statistical fluctuation, likely caused by the small 
number of sales events for any given customer.  There is little evidence for 
``black sheep'' with consistently low AUC scores, nor are there visible clusters.

\subsubsection{Model comparison}
\label{kdd:reco:compare}

While switching from training to validation set customers hardly affects
recommendation performance, there is a clear difference between the 
three fDNA models introduced in Section~\ref{kdd:network:architecture} when
the generalization to validation set articles is examined instead
(Table~\ref{kdd:reco:auc}).  Although attribute- and image-based fDNA perform 
at nearly equal levels in the training SKU set, the image-based model is 
distinctly inferior for articles in the validation set.  
Although this observation suggests that mere visual 
similarity is a worse predictor for customer interest than matching 
attributes, it might also involuntarily result from the widespread use of
article attributes in search filters and recommendation algorithms that
prejudice choice when customers browse the online shop.

\begin{figure}[t]
\centering
\includegraphics[width=\columnwidth]{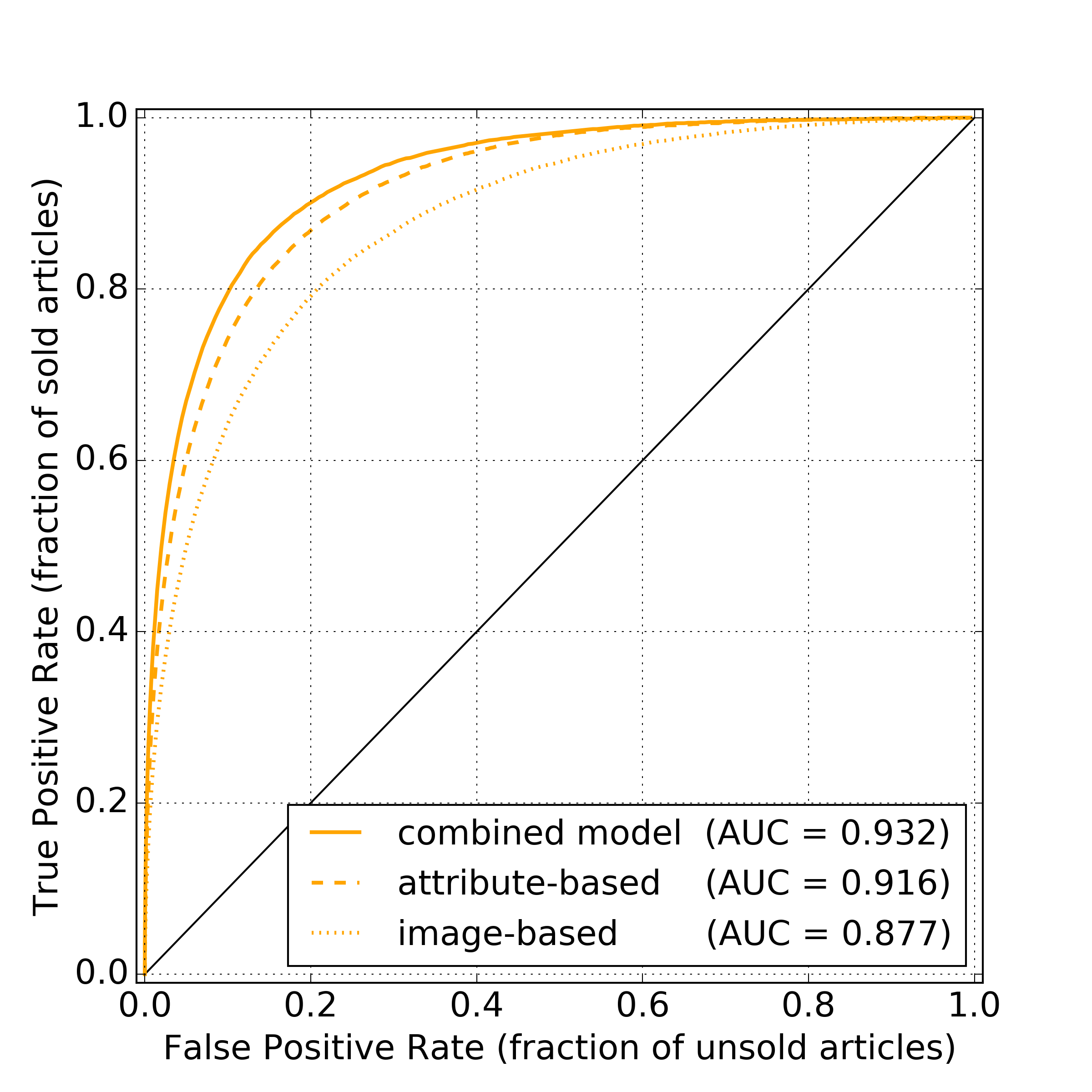}
\caption{Recommendation strength of different Fashion DNA models,
evaluated for SKU--customer validation data $\Pi^\mathrm{vv}$ 
(orange sector in Figure~\ref{kdd:network:maths:fig1}). The
attribute-based model (dashed curve) outperforms image-based fDNA (dotted curve). 
Combining attributes and images yields a superior fDNA model (solid line). 
The black diagonal line represents random guess performance.}
\label{kdd:reco:roc:fig3}
\end{figure}

As we are particularly interested in the ability of the models to generalize 
to unseen items and validation set customers, we calculated ROC curves from
the probability forecasts for the segment $\Pi^\mathbf{vv}$ 
(orange sector in Figure~\ref{kdd:network:maths:fig1}).  They are shown in 
Figure~\ref{kdd:reco:roc:fig3}.
The plot reveals that a ranking based on only six attributes (dashed curve)
beats recommendations based on the much richer image fDNA (dotted curve). 
As possible causes behind this observation, we note that items with very 
different function may look strikingly similar 
(e.~g., earmuffs and headphones), and that minor details
may alter the association of an article with a customer group (for instance,
men's and women's sport shoes often look much alike). As mentioned,
users commonly select attributes as filter settings for the 
Zalando online catalog, so the better performance of attribute fDNA may also
reflect the shopping habits of our customers. Importantly, we point out that integrating
attribute and image information boosts recommendation performance considerably,
as the ROC curve for the combined model fDNA (solid curve) shows.  
This indicates that images and attributes carry complementary information.

\begin{figure}
\centering
\includegraphics[width=\columnwidth]{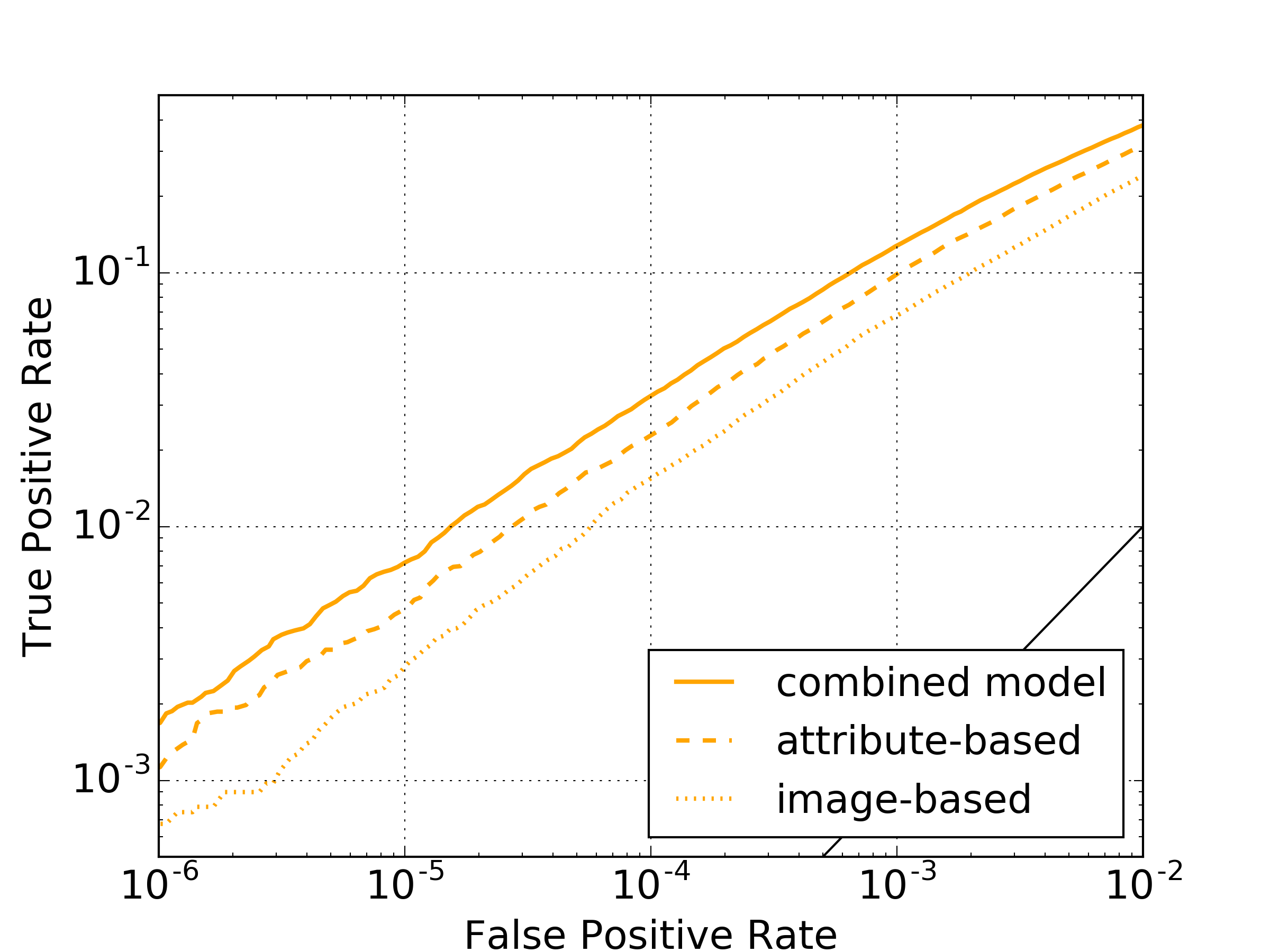}
\caption{Recommendation performance in the small sample size limit.
The logarithmic ROC plot is a detailed view of the lower left corner 
in Figure~\ref{kdd:reco:roc:fig3}.}
\label{kdd:reco:roc:fig4}
\end{figure}

For a more practical interpretation of the ROC analysis, we turn to
Figure~\ref{kdd:reco:roc:fig4}, which provides a magnified view of the ROC 
curves near the origin, i.e., for small true and false positive rates.  We
note that the ``positive'' (articles purchased) and ``negatives'' classes
in this case are extremely unbalanced (the average customer only acquires 
0.01\%\ of the article set). Hence, to a very good approximation, {\sl all}
120k articles in the validation set are negatives, and the rate of false positives
shown on the abscissa is essentially equivalent to the ratio of the number 
of recommended articles to the validation set size. (For instance, a false positive
rate of $10^{-4}$ represents the top-12 recommended articles.)
From the figure, we infer that the top recommendation represents about 
0.8\%\ of validation set purchases in the combined fDNA model --- for the average 
validation customer buying 17 items in the validation SKU set, this translates to a 
13\%\ ``success rate'' of the recommendation.
For top-12 (top-120) recommendation, one captures 3.5\%\ (13\%) of the items 
bought by shoppers similar to the ones in the customer training set.
(Of course, this ex-post-facto analysis cannot predict to which extent a
customer would have bought the top-rated items, had they been suggested to 
her as recommendations. This question properly belongs to the domain of 
A/B testing.)

\subsection{A Case Study}
\label{kdd:reco:case}

For illustration, we now present the leading recommendations of validation set articles 
for a sample frequent customer, a member of the training set, and compare results 
from the three Fashion DNA models introduced earlier.  (For lack of space, we only 
discuss a single case here. Our observations hold quite generally, though.)

\begin{figure*}
\centering
\includegraphics[width=\textwidth]{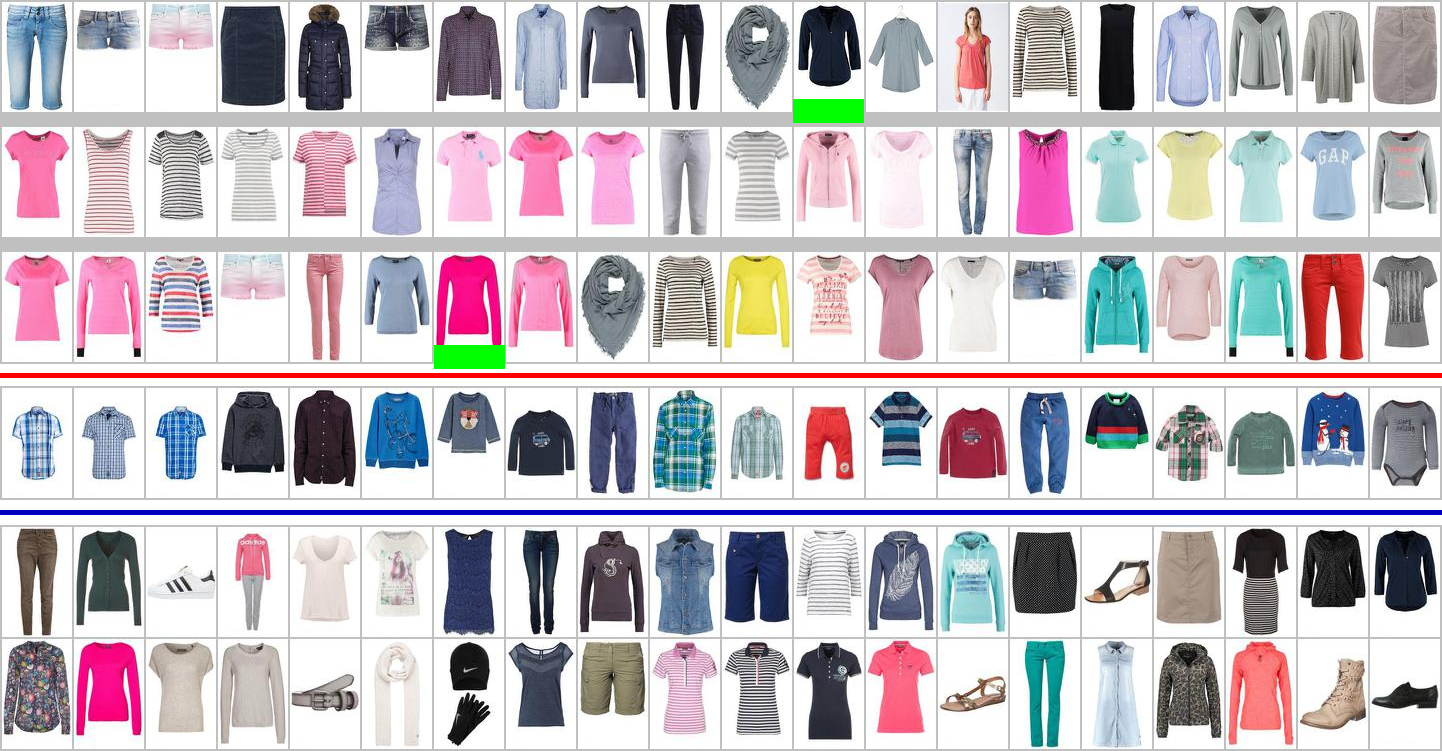}
\caption{Validation set recommendations for a frequent customer. Top:  
Leading 20 recommendations from attribute-based fDNA (first
row), image-based fDNA (second row), and combined fDNA (third row).  The
fourth row displays items judged least likely to be purchased in the
combined model.  Actual consumer purchases are displayed in the
row pair at the bottom.  Two matches are highlighted in green.}
\label{kdd:reco:roc:fig7}
\end{figure*}

Figure~\ref{kdd:reco:roc:fig7} displays the top-20 recommended fashion items
for the attribute-based model (top row), image-based model (second from top), 
and the combination model, our strongest contender (third row). The 
estimated likelihood of purchase decreases from left to right, with the first
choice exceeding 10\%; the mean forecast probability of sale for the
articles on display hovers around 5\%.  For contrast, we also display the items 
deemed {\sl least} likely in the combination model (fourth row). There, the
model estimates a negligible chance of sale (around $10^{-8}\,$\%).  Actual
consumer purchases are on view in the bottom rows.

Note that the items suggested by attribute fDNA, due to the lack of visual
information in training, appear quite heterogeneous.  The underlying similarity,
typically matching brands, remains concealed in the image. (Among the attributes
in Table~\ref{kdd:network:tags}, the brand and commodity group tags claim the 
lion's share of information, and major brands commonly try to gain market 
share by covering the whole fashion universe, from shoes to accessories.)
Although quite successful for our sample customer (one item highlighted in green 
was bought, and the top--100 recommendations capture four sales), the 
selection does not reflect her predilection with red and pink hues.
Unsurprisingly, image-based fDNA yields a visually more uniform result, often
aggregating items with a very similar look, here striped and pink shirts.
Although pleasing to the eye, the approach is less successful.
(None of the items shown was acquired, and there was a single ``hit'' in
the top 100 suggestions.)  The combination model integrates the
hidden features from the attribute model, and the visual trends (note that
recommended SKUs from both models are taken over), and offers a compact, yet more
varied selection.  In line with the general observation, it is also the most
performant model in our case study, with one item shown bought, and five
within the top 100 selection.  We finally remark that the least favorable
items (bottom row) are by themselves a coherent group (boys' clothes)
that summarizes qualities this customer is {\sl not} interested in: kids'
and men's apparel, plaid patterns.  This suggests that the models' negative 
recommendations may be actionable as well.

\section{Exploring article similarity}
\label{kdd:similarity}

As laid out in Section~\ref{kdd:network}, a central tenet of our approach 
is the assignment of articles $i$ to vectors $\mathbf f_i$ in a Fashion DNA space.  
Any metric in this space naturally defines a distance measure $D_{ik}$ 
between pairs of fashion items.  Cosine similarity is a simple choice 
that works well in practice:
\begin{equation}
D_{ik} \,=\, 1 - \frac{\mathbf f_i \cdot \mathbf f_k}{\sqrt{|\mathbf f_i|^2 |\mathbf f_k|^2}} \,.
\label{kdd:similarity:cosine}
\end{equation}
Every fDNA model gives rise to its own geometrical structure that emphasizes
different aspects of similarity.

\subsection{Next Neighbor Items}
\label{kdd:similarity:nn}

\begin{figure*}
\centering
\includegraphics[width=\textwidth]{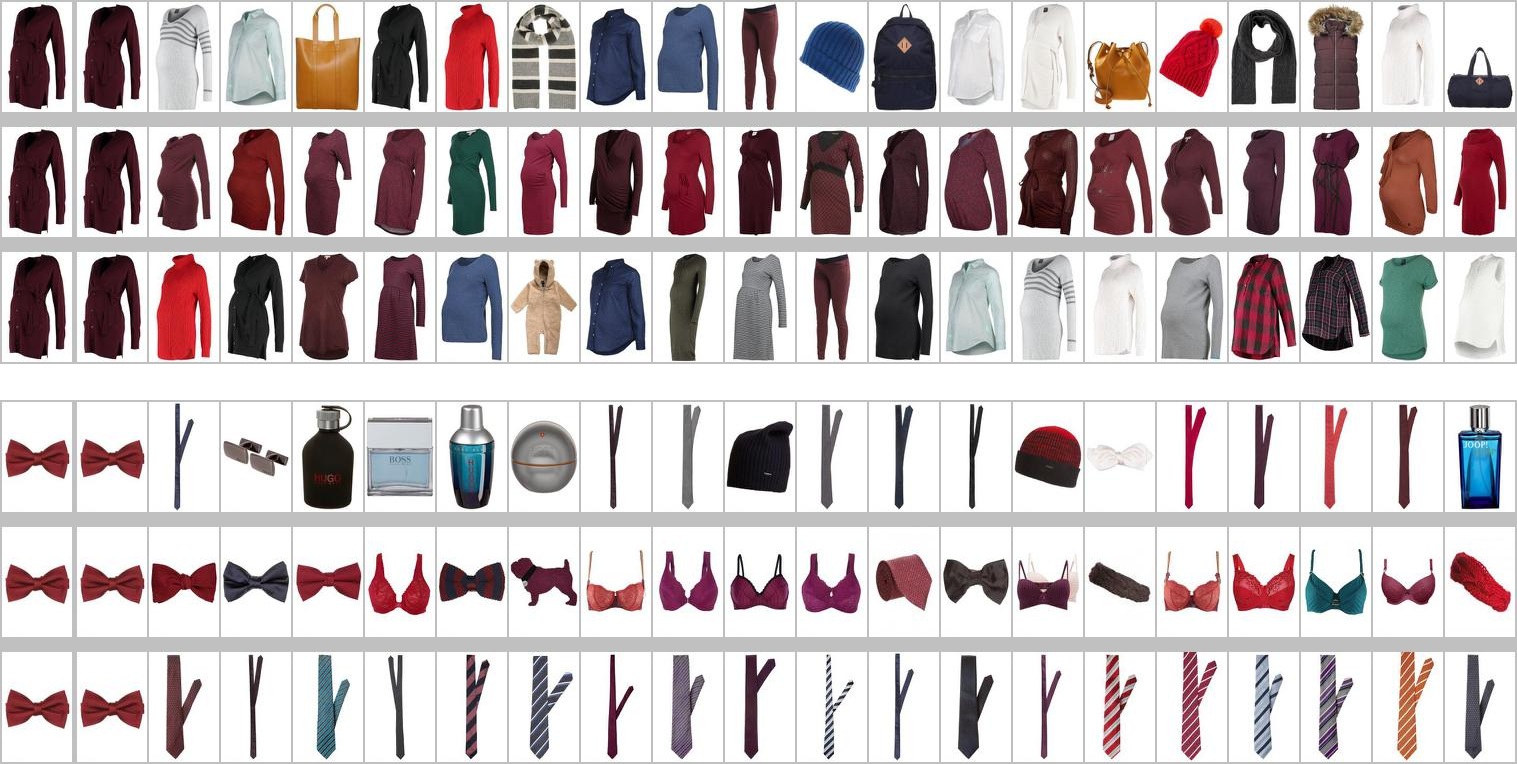}
\caption{Next-neighbor items in the 1.33m article catalog. The two
panels show next neighbors to the source items (maternity dress, bow tie) 
in the extreme left column, with cosine similarity decreasing to the right. 
Top rows:  Attribute-based fDNA, center rows: image-based fDNA, 
bottom rows: combined model.}
\label{kdd:similarity:nn:fig1}
\end{figure*}

We start out exploring local neighborhoods in the fDNA spaces.  Picking a
sample article (here, a maternity dress and a bow tie), we determine 
their angular distances (\ref{kdd:similarity:cosine}) to the standard SKUs, 
and select the nearest neighbors according to the three fDNA models..
They are displayed in Figure~\ref{kdd:similarity:nn:fig1}.

For both SKUs, attribute-based fDNA yields a heterogeneous mixture of articles
that hides the underlying abstract similarity:  The nearest neighbors to the 
dress all share the same brand, and in the case of the bow tie, men's accessories 
offered by various luxury brands are found.  For the image-based fDNA, visual
similarity is paramount; the search identifies many dresses aligned in color
and style, photographed in a similar fashion. The same analysis for the bow tie
reveals the risks underlying the visual strategy:  Here, the algorithm picks
completely unrelated objects that superficially look similar, like bras and a
dog pendant.  In either case, the most sensible results are returned by the 
combined attribute--image approach.  For the dress, the nearby items
are generally maternity attire that matches the original brand (note that the
mismatched objects in the attribute-based selection are now filtered out).
In the case of the bow tie, regular ties are identified as nearest neighbors:
The algorithm has learned that ties and bow ties share the same role
in fashion.  We also point out that the combined fDNA has included a baby romper
(of the same brand) in the neighborhood of the maternity dress -- a clear sign
that information propagates backward from the customer--item sales matrix into
the fDNA model.  Such conclusions are also inescapable when the global
distribution of items in the fDNA space is studied.

\subsection{Large-Scale Structure}
\label{kdd:similarity:tsne}

While the number of fashion items offered by Zalando is huge from a human 
perspective, they occupy the 256--dimensional fDNA space only sparsely.  
In order to visualize the arrangement of SKUs on a larger scale, we resort
to dimensionality reduction techniques, and find that t--SNE (stochastic
neighborhood embedding) is a suitable tool \cite{kdd:tsne} to reveal 
the structure hidden in the data.  The resulting maps are rather fascinating
descriptions of a fashion landscape that combines hierarchical order at several 
levels with smooth transitions.  

\begin{figure*}[t]
\centering
\includegraphics[width=\textwidth]{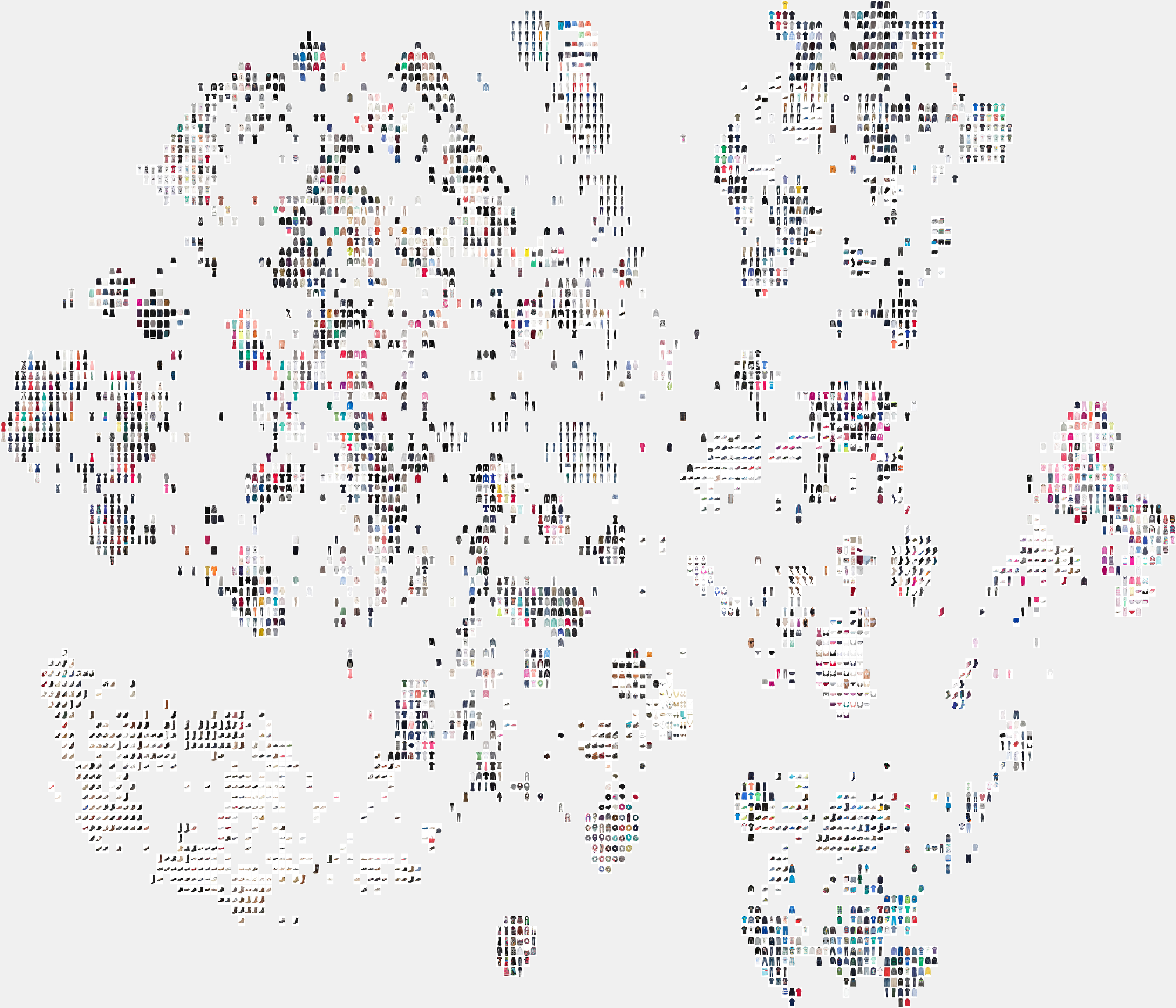}
\caption{t-SNE map of 4,096 fashion articles picked at random from the catalog, 
using the combined fDNA model.  Zoomable high--resolution view available 
online at \texttt{https://github.com/cbracher69/fashion-tsne-map}.}
\label{kdd:similarity:tsne:fig1}
\end{figure*}

\begin{figure}
\centering
\includegraphics[width=\columnwidth]{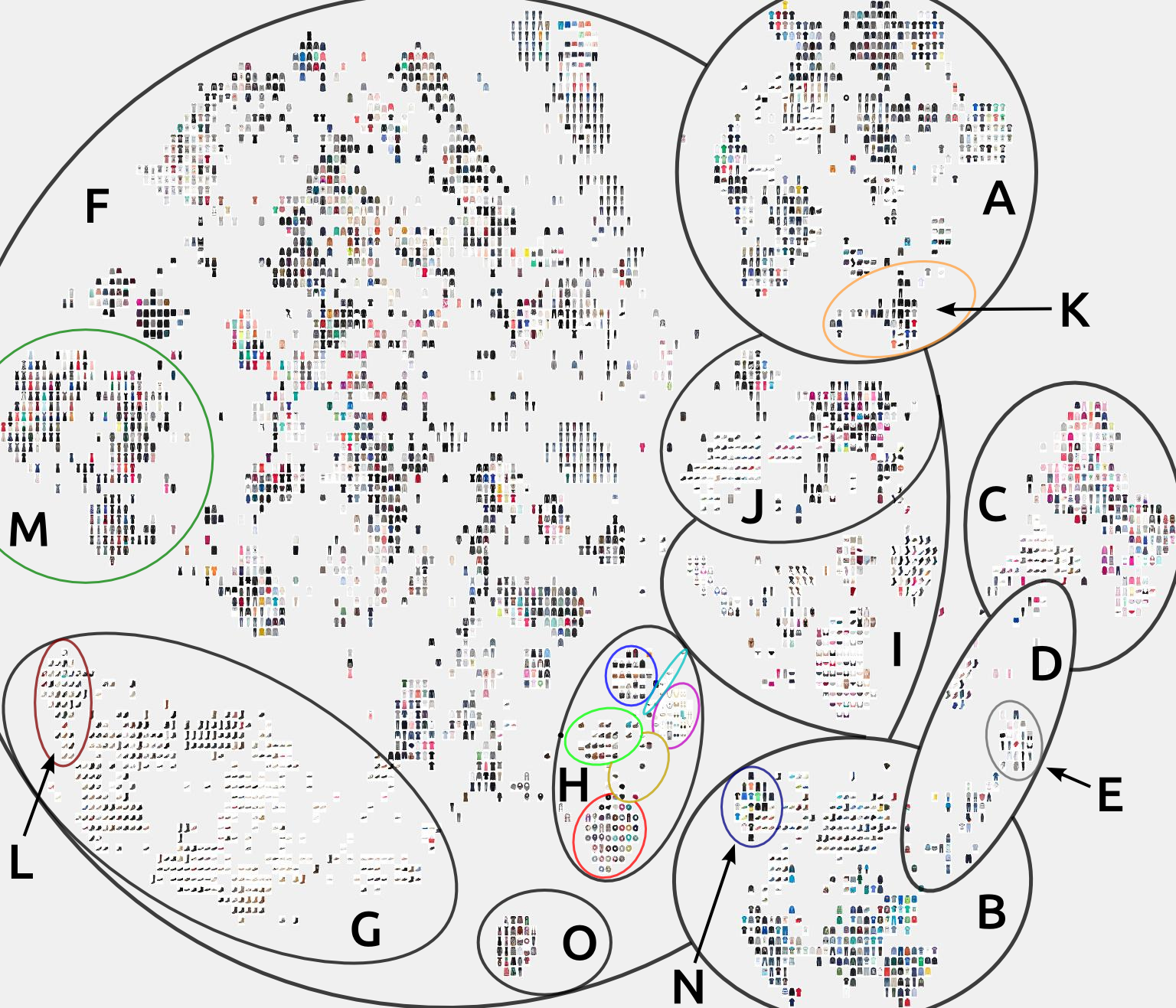}
\caption{A guide to highlights in the t-SNE fashion landscape.}
\label{kdd:similarity:tsne:fig2}
\end{figure}

Figure~\ref{kdd:similarity:tsne:fig1} displays such a map, generated from
4,096 randomly drawn articles from the catalog (subject to the weak restriction 
that the articles be sold at least 10 times to our 100,000 most frequent customers).
As underlying model, we used the combined attribute--image fDNA.
A glance at the map already shows the presence of orderly clusters, but
much interesting detail is revealed only when studying the arrangement in 
close-up view.  (A high-resolution image of the t-SNE map is posted online.)
In Figure~\ref{kdd:similarity:tsne:fig2}, we provide a guide to some 
``landmarks'' we found examining the image.

On the largest scale, the map is divided into three ``continents'':  On the
upper right, men's fashion items are assembled (\textbf{A}).  
The structure to the lower right contains kids' apparel and shoes, neatly 
separated into boys' (\textbf{B}) and girls' fashion (\textbf{C}), 
connected by a narrow isthmus of baby (\textbf{D}) and maternity items
(\textbf{E}).  The huge remainder of the map (\textbf{F}) is devoted to 
womens' fashion gear, with a core body of assorted clothing (upper left), 
surrounded by satellites of casual and fashion shoes (\textbf{G}, lower left),
accessories (\textbf{H}), lingerie, hosiery, and swimwear (\textbf{I}), and 
sport shoes and gear (\textbf{J}).  Each of the regions is further
subdivided (e.~g., accessories (\textbf{H}) are separated into scarves 
(red), belts (green), hats (brown), bags (blue), jewelry (purple), 
and sunglasses (cyan)).  It is remarkable that the separation at the top
level is essentially complete, as there is for instance no mixing between
male and female items.  Instead, subcategories are replicated in each
section (such as a male sports cluster (\textbf{K}) that, naturally enough, 
faces the much larger female counterpart (\textbf{J})).  But there is also 
gradual progression in some structures:  For instance, heel heights increase 
as one travels clockwise along the outer rim of the shoe cluster 
(with high heels (\textbf{L}) being opposite dresses (\textbf{M}) on the 
apparel side).  Another example is kids' clothing, where size and age group 
increases as one travels away from the maternity ``center'' (\textbf{E}).  
It is notable that at the lower levels of organization,
clustering mostly occurs by shape or pattern, and only rarely by function
(e.~g., there is a ``soccer cluster'' (\textbf{N}) in the boys' section), 
or brand loyalty, the {\sl Desigual} ``island'' (\textbf{O}) at the very 
bottom being the only conspicuous example.

\section{Conclusions}
\label{kdd:outlook}

In this paper, we introduced the concept of Fashion DNA, a mapping of
fashion articles to vectors using deep neural networks that is optimized to 
forecast purchases across a large group of customers.
Being based on item properties, our approach is able to circumvent the cold 
start problem and provide article recommendations even in the absence of
prior purchase records.  Likewise, the model is flexible enough to generate
sales probability predictions of comparable quality for validation customers.  
We demonstrated that an fDNA model based on article attributes and images 
generalizes well and suggests relevant items to shoppers.  
The combination of article and sales information imprints a wealth of 
structure onto the item distribution in fDNA space.

We plan to enrich the model with additional types of fashion-related data, such as
ratings, reviews, sentiment, and social media, which will require extensions of the 
deep network handling the information, e.~g.\ natural language processing
capability.  With an increasing number of information channels, for many items
only partial data will be available. To render the network resilient to missing 
information, we will further experiment with drop-out layers in our architectures.

An important aspect currently absent in our model is the temporal order of sales 
events.  Customer interest and item relevance evolve over time, whether by season, 
by fashion trends, or by personal circumstances.  Capturing and forecasting such 
variations is a difficult task, but also a valuable business proposition that may 
be tackled by introducing long short-term memory (LSTM) elements into our network.

\section{Acknowledgments}
We would like to thank Urs Bergmann for valuable comments on this paper.


\end{document}